\documentclass[10pt,a4paper,twoside]{article}
\usepackage{epsfig}
\usepackage{baltlat6}
\usepackage{array}
\usepackage{amssymb}
\begin{document}
\ \
\titlehead{Baltic Astronomy, vol.18, 293-296, 2009}

\titleb{IMPACT ON COSMOLOGY OF THE CELESTIAL ANISOTROPY OF THE
SHORT GAMMA-RAY BURSTS}

\begin{authorl}
\authorb{A. M\'esz\'aros}{1},
\authorb{L.G. Bal\'azs}{2},
\authorb{Z. Bagoly}{3} and
\authorb{P. Veres}{3,4}
\end{authorl}

\begin{addressl}
\addressb{1}{Charles University,
Faculty of Mathematics and Physics,
Astronomical Institute,
V Hole\v{s}ovi\v{c}k\'ach 2, 180 00 Prague 8,
Czech Republic;
meszaros@cesnet.cz}
\addressb{2}{Konkoly Observatory, PO BOX 67, H-1525 Budapest, 
Hungary; balazs@konkoly.hu}
\addressb{3}{Lab. for Information Technology, E\"{o}tv\"{o}s University, 
P\'azm\'any P. s. 1/A, H-1518 
Budapest, Hungary; zsolt@yela.elte.hu; veresp@elte.hu}
\addressb{4}{Dept. of Physics, Bolyai Military University, H-1581 
Budapest, POB 15, Hungary}
\end{addressl}

\submitb{Received: 2009 July 20; accepted: 2009 December 1}

\begin{summary} 
Recently the anisotropy of the short gamma-ray bursts detected by BATSE
was announced (Vavrek et al. 2008). The impact of this discovery 
on cosmology is discussed. It is shown that the anisotropy found may cause 
the breakdown of the cosmological principle.
\end{summary}


\resthead{Impact on cosmology of the anisotropy of short GRBs}
{A. M\'esz\'aros, L.G. Bal\'azs, Z. Bagoly, P. Veres}

\sectionb{1}{INTRODUCTION}

Several statistical studies show that there are short, intermediate and
long gamma-ray bursts (GRBs) 
(Horv\'ath 1998, 2002, 2009; Mukherjee et al. 1998; 
Hakkila et al. 2000, 2004; Horv\'ath et al. 2002, 2004, 2008; 
Bal\'azs et al. 2004; Gehrels et 
al. 2006; \v{R}\'{\i}pa et al. 2009; 
Huja et al. 2009). The names of groups follow from the facts that 
they can be separated mainly with respect to their durations. 
The short and long GRBs are associated with different objects (Bal\'azs et 
al. 2003; Nakar 2007) but the interpretation of
the intermediate GRBs is still unknown (Horv\'ath et al. 2008).

The sky distributions of GRBs of different durations are shown in Figure 1.
The short GRBs, collected in the BATSE catalog (Mallozzi et 
al. 2001), are distributed anisotropically. The sky distribution of 
the intermediate GRBs is also anisotropic. 
No obvious anisotropy was found for the long bursts - hence they can be 
distributed isotropically (Vavrek et al. 2008). 

In this paper we discuss cosmological consequences of 
anisotropic distributions. Obviously,
once the cosmological principle is fulfilled, i.e., the 
mass distribution in the Universe is homogeneous and isotropic on a large 
scale (Peebles 1993), then isotropical distribution of GRBs is expected.
This seems to be valid for the long GRBs, but not for the 
remaining two types. 
This may mean that in the range of redshifts, where these two 
types of GRBs happen, the cosmological principle is not valid.
In this paper we try to verify this possibility by estimating  
redshifts of the short GRBs from the BATSE catalog. The  
redshifts of the intermediate bursts are beyond the scope of this article.

\begin{figure}[!tH]
\vbox{
\centerline{\psfig{figure=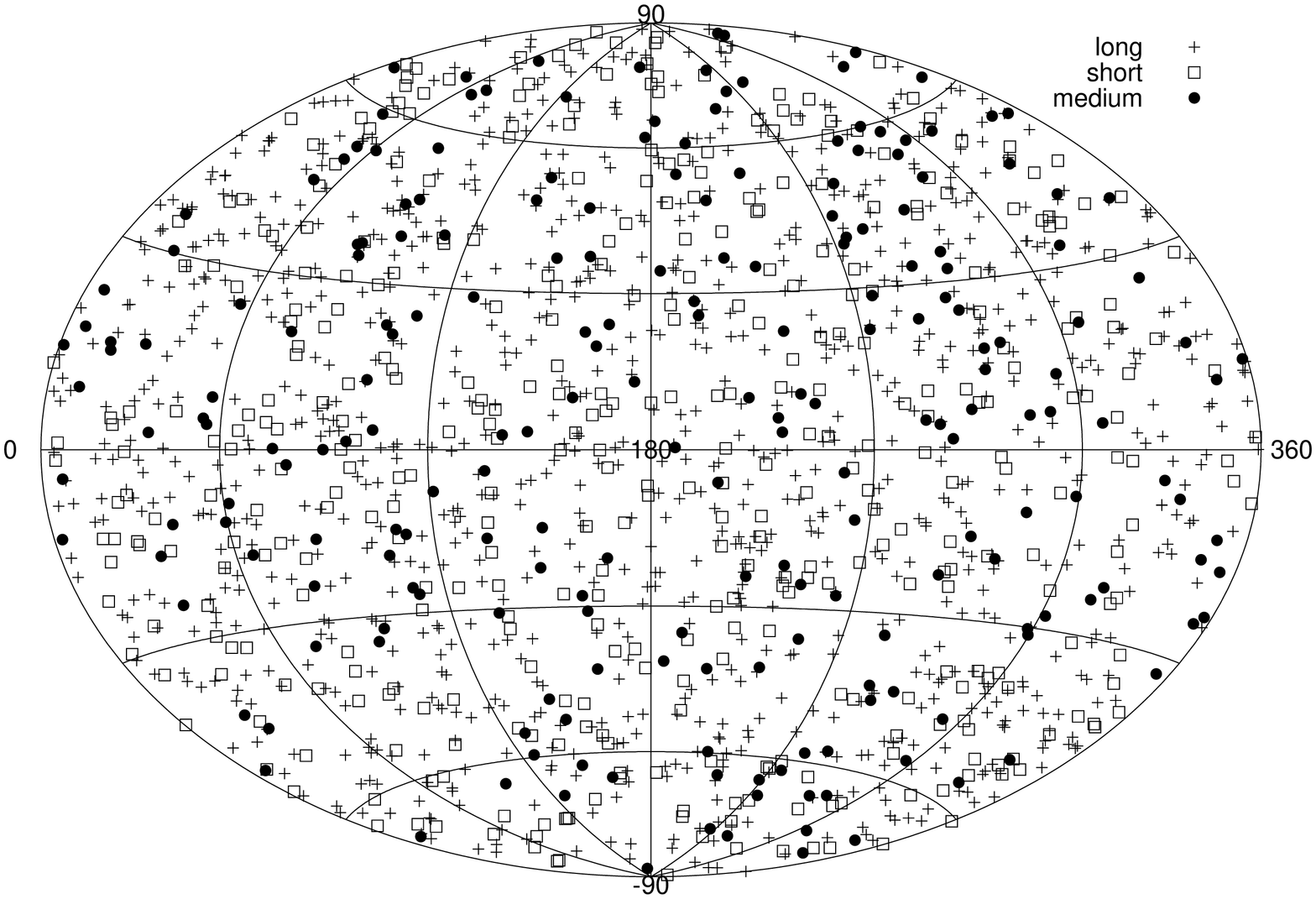,width=88mm,angle=-90,clip=}}
\vspace{1mm}
\captionb{1}
{Sky distribution of the BATSE bursts. The three different subgroups are
denoted as explained in the insert ('medium' means intermediate GRBs).
}
}
\end{figure}

\begin{table}
\caption{Short GRBs from Swift with known redshifts.}
\begin{center}
 \begin{tabular}{lcccc}
              \hline
GRB & Redshift    & $T_{90}$   & Fluence      & Peak-flux   \\
name &     $z$      &  s     & $10^{-7}$erg/cm$^2$   & ph/(cm$^2$s)  \\
\hline
050509B  & 0.22 & 0.073 & 0.09 &  0.28 \\
050813  & 1.80 & 0.44 &  0.44 & 0.94\\
051221A & 0.55 &1.4 &11.5& 12.0\\
060502B  & 0.29 & 0.13 & 0.4 & 0.62\\
061201  & 0.11 & 0.76 & 3.34 & 3.86 \\
061217  & 0.87 & 0.212 & 0.42  &1.49 \\
070429B  & 0.90 & 0.47 & 0.63 & 1.76 \\
070724A & 0.46 & 0.4 & 0.3 & 1.0\\
071227  & 0.38 & 1.8 & 2.2 & 1.6\\
\hline 
\end{tabular}
\end{center}
\end{table}

\sectionb{2}{REDSHIFTS OF THE SWIFT SHORT BURSTS}

No directly measured redshifts are known for the BATSE short bursts.
Therefore we will try to estimate the range of their redshifts indirectly,
comparing them with known redshifts of short bursts from the Swift 
database. Short GRBs from BATSE will be discussed in the next section.

In the Swift database till 2009 February 28 there were 133 GRBs with 
known redshifts (Gehrels et al. 2009). 
Among them we find only nine (Table 1) which are considered as short by
Zhang et al. (2009) and for which the measured
$T_{90}$ values (in seconds) are not higher than
$2(1+z)$ ($z$ is the redshift). 

Note that there is a controversy concerning GRB080913 with
$z = 6.7$. It has $T_{90} =  8\; s$, and thus it can still be a short 
burst, because its intrinsic duration is smaller than $2\, s$. 
According to Zhang et al. (2009) this GRB probably is not a 
short burst, although such a possibility is not excluded. We decided
not to include GRB 080913 in the list of short GRBs.

The mean of the
redshifts of GRBs from Table 1 is $\mu_{z} = 0.62$, the dispersion
is $\sigma_{z} = 0.49$, and the median is $z = 0.46$.
The redshifts cover the interval between $0.11$ and $1.8$.
 
\sectionb{3}{THE REDSHIFTS OF BATSE'S SHORT BURSTS}

The short GRBs with known redshifts detected by 
Swift and the short GRBs 
from the BATSE catalog (Mallozzi et al. 2001) are not necessary located
in the same redshift range. For example, due to 
some selection effects (the Swift data and the BATSE data were 
obtained by different instruments), we cannot exclude the
possibility that the short GRBs in the BATSE sample are predominantly
at $z <0.1$. If this were the case, then their anisotropy would be expected, 
because in the Universe bold inhomogeneities are observed up to  $z = 0.1$ 
(Peebles 1993).

A comparison of the GRBs from Table 1, and the 406 short GRBs from 
the BATSE catalog shows that this is not the case. 
Bal\'azs et al. (2003) have shown that
the mean fluence of the BATSE sample is $5 \times 10^{-7}$ erg/cm$^2$ with a
dispersion as large as $\sim 50 \%$. Table 1 shows that the mean
fluence of the Swift sample is $2.1 \times 10^{-7}$ erg/cm$^2$, again with 
a large $\sim 100 \%$ dispersion. The two fluences are
in a good agreement. The mean peak-flux for the BATSE sample is
$3$ photons/(cm$^2s)$ (Bal\'azs et al. 2003) and for
the Swift sample from Table 1 it is $2.4$ photons/(cm$^2s)$. Bearing in 
mind that 
in both cases the dispersions are large ($\sim (50-100)\%$), the peak
fluxes are also in good agreement. All this suggests that the redshifts of 
short GRBs collected both from the BATSE and Swift catalogs are mostly 
between $0.1$ and $0.9$.

Two notes are essential here. First, Nakar (2007) considers that
the short/hard GRBs should be formed in the mentioned redshift range. 
Second, we cannot exclude that among the short GRBs even higher
redshifts ($z > 1$) can be present, if 
GRB080913 belongs to the short GRBs (Zhang et al. 2009). Even GRB090423 
with the highest known redshift ($z = 8.2$) can happen to belong to this
type (Krimm et al. 2009). 

\sectionb{4}{CONCLUSION}

The cosmological principle requires that the Universe must be
spatially homogeneous and isotropic on the scales larger than the size of 
any structure (void, filament, supercluster, etc.). In other words, the
scale of averaging must be higher than the size of any structure. 
But, at the same time, the averaging scale must be smaller than 
the Hubble radius (Peebles 1993). Most of the short GRBs which are 
distributed anisotropically are at redshifts up to $0.9$.
The proper-motion distance (Weinberg 1972) corresponding to 
$z = 0.9$ is $3$ Gpc for the Hubble radius $14$ Gpc. 
The proper motion distance $d_{PM}$ is defined as  
$d_{PM} = d_{LM}/(1+z)$, where $d_{LM}$ is the luminosity distance 
(Weinberg 1972). All distances here are taken as proper-motion distances 
for the most probable 
cosmological parameters  $H_o = 71$ km/(sMpc), $\Omega_M = 0.27$ and 
$\Omega_{\Lambda} = \lambda c^2/(3H_o^2) = 0.73$; $H_o$ is 
the Hubble-constant, $\lambda$ is the cosmological constant, $c$ is the 
velocity of light in vacuum, and $\Omega_M$ is the ratio of density of 
the Universe to the critical density (Wright 2009).
 
Thus, the short GRBs suggest the presence of structures of the Gpc 
scales. This is a 
great challenge to cosmology, because a scale of averaging between 
3 and 14 Gpc should be present; in principle, this is still possible, but 
obviously complicated enough. If there are short GRBs also at $z > 0.9$, 
then any averaging is already impossible, because  
the corresponding proper-motion distance for $z=8.2$ is $9.2$ Gpc; i.e. 
$\simeq 66 \%$ of the Hubble radius.
   
\bigskip

ACKNOWLEDGMENTS.
The useful discussions with J. \v{R}\'{\i}pa are acknow\-ledged.
This study was supported by the GAUK grant No. 46307, by the OTKA
grant K077795,
and by the Research Program MSM0021620860 of the Ministry
of Education of the Czech Republic.

\References

\refb Bal\'azs L.G., Bagoly Z., Horv\'ath I. et al. 2003, A\&A, 401, 
129

\refb Bal\'azs L.G., Bagoly Z., Horv\'ath I. et al. 2004, Baltic 
Astronomy, 13, 207

\refb Gehrels N., Norris J.P., Barthelmy S.D. et al. 2008, Nature, 444, 1044

\refb Gehrels N., Newman P., Myers J.D. et al. 2009, Swift database, \\
\texttt{http://heasarc.gsfc.nasa.gov/docs/swift/swiftsc.html}

\refb Hakkila J., Giblin T.W., Roiger R.J. et al. 2004, Baltic 
Astronomy, 13, 211

\refb Hakkila J., Haglin D.J., Pendleton G.N. et al. 2000, ApJ, 538, 165

\refb Horv\'ath I. 1998, ApJ, 508, 757

\refb Horv\'ath I. 2002, A\&A, 392, 791

\refb Horv\'ath I., M\'esz\'aros A., Bal\'azs L.G., Bagoly Z. 2004, Baltic 
Astronomy, 13, 217

\refb Horv\'ath I., Bal\'azs L.G., Bagoly Z. et al. 2006, A\&A, 447, 23

\refb Horv\'ath I., Bal\'azs L.G., Bagoly Z., Veres P. 2008, A\&A, 489, L1

\refb Horv\'ath I. 2009, Ap\&SS, 323, 63

\refb Huja D., M\'esz\'aros A., \v{R}\'{\i}pa J. 2009, A\&A, 504, 67

\refb Krimm H.A., Norris J.P., Ukwatta T.N. et al. 2009, GCN Circular 9241
 
\refb Mallozzi R. S., Fishman G., 
Connaughton V. et al. 2001, Current BATSE Gamma-Ray Burst Catalog,
\texttt{http://gammaray.msfc.nasa.gov/batse/grb/catalog}

\refb Mukherjee S., Feigelson E.D., Jogesh Babu G. et al. 
1998, ApJ, 508, 314

\refb Nakar E. 2007, Phys. Rep., 442, 166

\refb Peebles P.J.E. 1993, {\it Principles of Physical Cosmology}, 
Princeton Univ. Press

\refb \v{R}\'{\i}pa J., M\'esz\'aros A., Wigger C. et al. 2009, A\&A, 498, 399

\refb Vavrek R., Bal\'azs L.G., M\'esz\'aros A. et al. 2008, MNRAS, 391, 1741

\refb Weinberg S. 1972, {\it Gravitation and Cosmology}, 
J. Wiley and Sons., 1972

\refb  Wright N. 2009, 
\texttt{http://www.astro.ucla.edu/$\sim$wright/cosmolog.htm}

\refb Zhang B., Zhang B.-B., Virgili F.J. et al. 2009, ApJ, 703, 1696 

\end{document}